\documentclass{WileyMSP-template}

\usepackage{lineno}
\usepackage{siunitx}
\usepackage{placeins}

\newcommand{\cedge}{\textit{coupling-edge }}
\usepackage[style=numeric,sorting=none]{biblatex}
\begin{document}

\pagestyle{fancy}
\rhead{\includegraphics[width=2.5cm]{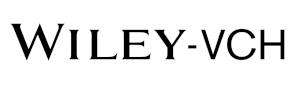}}

\title{Photonic neuromorphic processing with coupled spiking silicon microrings}

\maketitle


\author{Giovanni Donati*}
\author{Stefano Biasi**}
\author{Lorenzo Pavesi**}
\author{Antonio Hurtado*}

\begin{affiliations}
*Institute of Photonics, Department of Physics, University of Strathclyde, Glasgow, United Kingdom\\
**Nanoscience Laboratory, Department of Physics, University of Trento, Trento, Italy

\end{affiliations}



\begin{abstract}
Understanding the physical computing mechanisms of individual network nodes is essential for scaling neuromorphic photonic architectures. This work proposes a compact passive nonlinear photonic core based on a Side-Coupled Integrated Spaced Sequence of Resonators (SCISSOR) made of three nominally equal microrings and investigate its computing capabilities. Its nonlinearities and internal feedback enable analogue, spiking, and bistable responses that are accessed by tuning the injection power and wavelength. Implemented as a single nonlinear node in a time-multiplexed reservoir computing, the SCISSOR achieves error-free classification on the Iris dataset and accuracies above 97$\%$ on the Sonar task, using both analogue and digital reservoir representations with 150 virtual nodes. In the digital scheme, spiking dynamics naturally generate sparse reservoir states, enabling efficient classification even with a single spike. Intriguingly, optimal operating points are at the boundaries where sharp transitions in dynamical complexity and/or output power occur. In these points, the SCISSOR supports high task-performance, opening novel strategies for future on-chip training. Spiking and thermal bistabilities also participate to enhance the computational performance at low injected powers below 4 mW. These results suggest optical coupled microring resonators as effective building blocks for future edge computing and neuromorphic photonic systems.
\end{abstract}

\section{Introduction}\label{sec_Intro}
With the approaching end of the Moore's law, the computing community is increasingly seeking alternative technologies to improve performance in processing the ever-growing amount of data. 
Neuromorphic photonics is widely regarded as a key enabling technology for next-generation information processing \cite{shastri2021photonics,schuman2022opportunities}. This applies particularly when the input information is natively optical and can be pre-processed locally by a photonic processor, avoiding repeated electro–optical conversions and minimizing latency. In such architectures, the photonic subsystem performs the nonlinear transformation and temporal processing of the input, while a low-complexity electronic readout extracts task-relevant information. This division of roles naturally aligns with edge computing scenarios, where speed and energy efficiency are critical \cite{cao2018edge}.
Among neuromorphic paradigms, Reservoir Computing (RC) \cite{maass2002real, jaeger2001echo} leverages the natural dynamics of physical systems to perform computation without requiring on-chip training of the internal connections and is therefore well suited for applications requiring fast inference, adaptive control, and real-time processing like edge computing.
There is a pressing need, therefore, to develop an improved understanding of the capabilities and integrability of photonic RC. 

Silicon photonics provides a scalable and cost-effective platform for integrated neuromorphic systems, benefiting from its CMOS compatibility, with microring resonators (MRRs) emerging as fundamental building blocks for neuromorphic computing \cite{biasi2024photonic}.
Their wavelength filtering properties enable linear optical mixing \cite{biasi2023array}, reconfigurable weighting \cite{tait2016microring, huang2021silicon}, and nonlinear activation \cite{ashtiani2022chip}, while high quality factor designs support strong light confinement which allows for two-photon absorption (TPA) physics at relatively low optical injection around 1550 nm \cite{borghi2017nonlinear}. 
TPA generates extra free carriers and heat, inducing refractive index changes via free-carrier dispersion (FCD) and thermo-optic effects (TOE) that evolve on distinct timescales (free carrier lifetime $\tau_{fc}$=1-45 ns, thermal relaxation time $\tau_{th}$=60-280 ns) \cite{borghi2021modeling}. The competition between these processes within the MRR, shifts accordingly its resonance position, hence its transmittance, giving rise to nonlinear behaviours, such as thermal bistability \cite{almeida2004optical} and self-pulsation dynamics \cite{van2012simplified,borghi2021modeling, xiang2022all}. 
Nonlinear transient dynamics based on free carriers and temperature effects have been successfully exploited in analogue RC implementations based on a single MRR node, where time-multiplexed virtual neurons are sampled from the analogue dynamics \cite{borghi2021reservoir, bazzanella2022microring, donati2021microring, donati2024time, giron2024effects, gretter2025dynamic}.
More recently, the spiking responses exhibited by single MRRs under pulsed excitation have enabled digital RC schemes, in which the reservoir states are reduced to binary values, either 0 or 1, forming sparse digital patterns that simplify the electronic post-processing \cite{donati2025all}. 
However, the limited optical dynamics achievable with a single MRR in digital RC constrains the variety of output patterns and, ultimately, the performance on the task. 
Increasing the computational capability therefore requires more complex photonic structures of interacting nonlinear elements \cite{beltramo2025injection, li2025nonlinear}.
Matrices of coupled MRRs designed to operate in nonlinear regimes triggered by TPA have been recently investigated in the context of analogue RC, both numerically \cite{denis2018all, ren2024photonic, dong2026deep} and experimentally with demonstrated designs scaled up to arrays of 8x8 MRR \cite{lugnan2025reservoir, foradori2025neuromorphic}. The latter showcase long-term thermal memory (tens of $\mu s$) based on self-pulsation dynamics \cite{biasi2024exploring}, and integrate emergent learning capabilities thanks to integrated phase-change materials \cite{lugnan2025emergent}. 
Yet, scaling photonic computing architectures in a ``black box" approach limits the efficiency of large-scale designs and makes the operational parameter space growing rapidly. This issue becomes even more critical when explicit weight control is required, since thermal crosstalk introduces strong interdependence among neighboring elements \cite{biasi2022effect, teofilovic2024thermal}. Understanding the computing capabilities of reduced structures, which still exhibit complex dynamics, is therefore a crucial step toward efficiently scaling up optical neuromorphic computing designs on-chip.

\begin{figure}[t]
\centering
\includegraphics[width=1\textwidth]{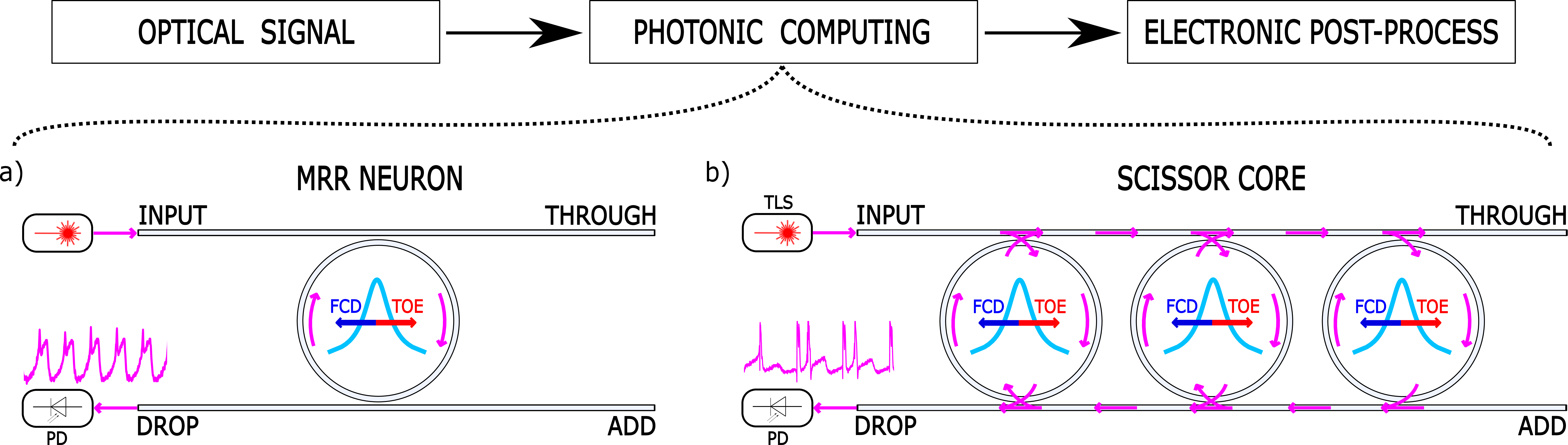}
\caption{Schematic of an integrated silicon-photonic SCISSOR neuromorphic core for edge computing applications. Two-photon absorption induces free-carrier dispersion (FCD) and thermo-optic effects (TOE), which together govern the nonlinear transformation applied to optical input signals, reducing the complexity of subsequent electronic post-processing.}\label{fig_intro}
\end{figure}

Here, we propose and experimentally investigate the potential of a Side-Coupled Integrated Spaced Sequence of Resonators (SCISSOR), consisting of three nominally identical self-pulsating silicon MRRs coupled between two bus waveguides (Fig. \ref{fig_intro}) as a photonic neuromorphic core for optical computing and RC. 
As predicted in \cite{mancinelli2014chaotic}, this structure represents a minimal photonic system capable of exhibiting aperiodic dynamics arising from multi-resonator TPA-based nonlinear interaction, enabling access to ``edge of chaos'' regimes which are known to often enhance the computational capability \cite{langton1990computation, carroll2020reservoir}.  
Here we show that when implementing the SCISSOR nonlinearity in both analogue and digital RC architectures, error-free classification is achieved on the Iris dataset and accuracy exceeding $97\%$ on the more challenging Sonar task. 
Notably, the SCISSOR spiking dynamics allows for more diverse yet sparse digital representations of the input information, making it promising for efficient photonic-electronic RC implementations.
Moreover, correlating the task performance with a detailed nonlinear characterization of the device highlights \cedge operating points, defined as regions where transitions in dynamical complexity and/or output optical power occur at the boundaries of weak and strong coupling regimes, which can support high computational performance. Although other optimal operating points exist, these boundary regions naturally constrain the operational space, offering strategies for training and control of future scaled-up designs. 

The remainder of the manuscript is organized as follows. Section \ref{sec_complexity} characterizes the nonlinear dynamics of the SCISSOR and identifies injection conditions associated with the loss of temporal correlation under continuous wave (CW) excitation. Section \ref{sec_RC} assesses the computational performance of the SCISSOR within analogue and digital RC frameworks using the Iris Flower and Sonar dataset classification tasks. The Sonar task, being more challenging, shows that \cedge regimes can provide favourable operating conditions for high and robust performance. Finally, Section \ref{sec_Conclusions} outlines the conclusions and future perspectives of this work.

\FloatBarrier
\section{SCISSOR nonlinear characterization}\label{sec_complexity}  

\begin{figure}[!t]
\centering
\includegraphics[width=1\textwidth]{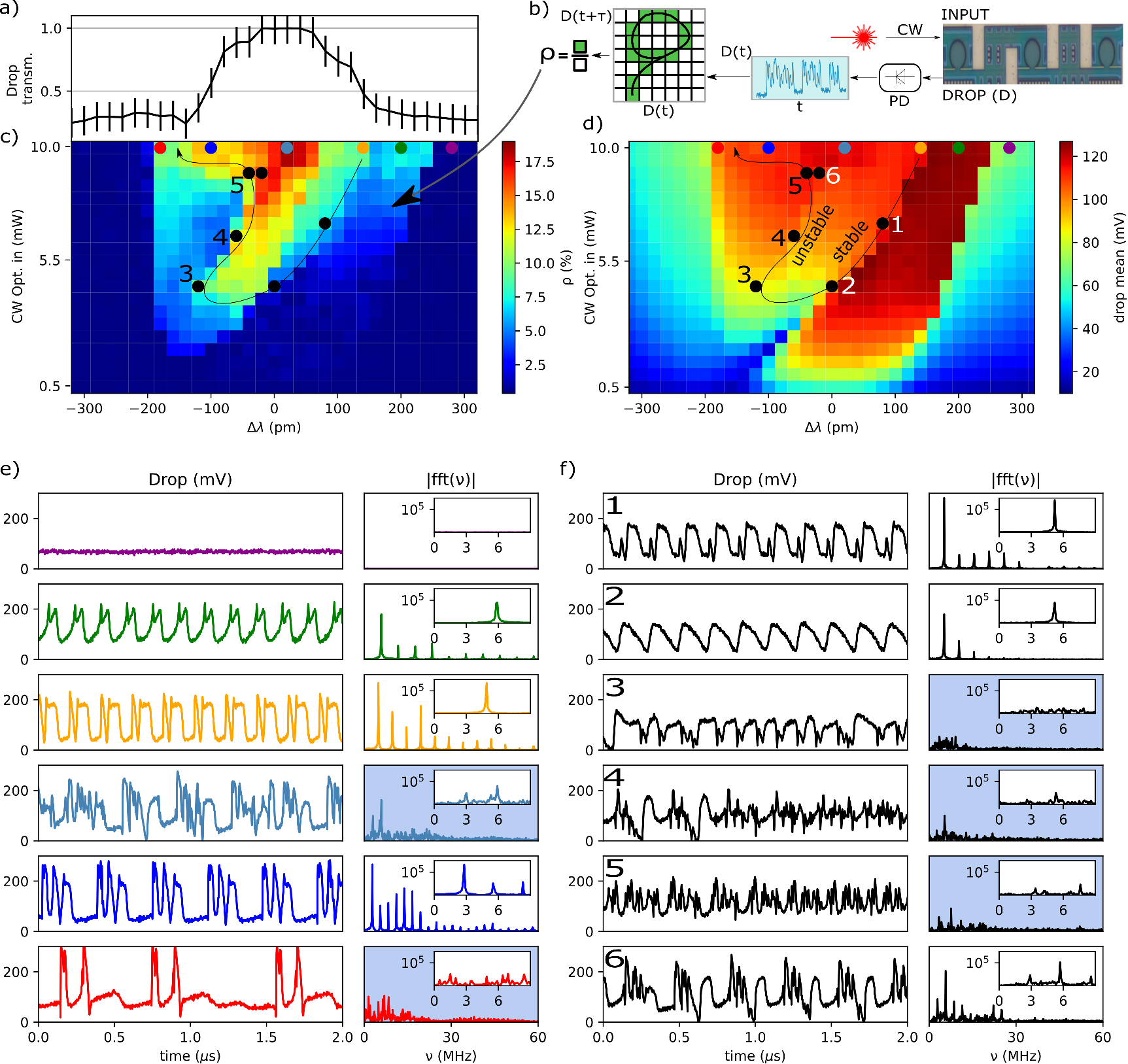}
\caption{Experimental characterization of a 3-ring SCISSOR under continuous wave (CW) optical injection. (a) Linear drop spectrum. (b) Experimental scheme where CW laser signal is injected at the input port of the device and the corresponding drop port is analysed to recreate a phase space relying on the Takens' theorem. (c) Estimation of the density of the phase space for multiple $\Delta\lambda$-CW Optical power injection configurations and, (d) corresponding mean value of the detected drop signal. (e-f) Examples of SCISSOR's dynamics and corresponding fast Fourier transforms (fft) under injection configurations labelled by coloured (e) and numbered (f) circles, in (c) and (d). The time traces and their fft are evaluated over 10~$\mu$s of recorded data with 2 $ns$ sample rate; for visualization, only a 2 $\mu$s portion of the temporal signal is shown in the panels. Fft panels showing the lost of harmonic peaks have a light-blue background colour. }\label{fig_complexity}
\end{figure}

The SCISSOR under study consists of three self-pulsating silicon MRRs with a radius of 7 $\mu$m, evanescently-side coupled by two bus waveguides with a symmetric gap of 180 nm. A center-to-center distance of 65.973 $\mu$m between the MRRs is chosen to satisfy internal constructive interference conditions \cite{smith2004coupled,xu2006experimental, mancinelli2011coupled}.
Figure \ref{fig_complexity}(a) shows the linear spectrum of the device, confirming that the three MRR resonances nearly overlap around $\lambda_0=1558.615$ nm. 
This overlap enables coherent optical feedback between the MRRs for injection wavelengths $\lambda_p$ coupled to the device.
Activating TPA in any of the MRRs perturbs the coherent feedback, giving rise to complex nonlinear responses \cite{mancinelli2014chaotic}.
The nonlinear dynamics supported by the SCISSOR, and therefore its potential as a neuromorphic processing core, are probed by injecting CW light under different configurations of wavelength detuning $\Delta\lambda=\lambda_p-\lambda_o$ and optical power.  
As shown in Fig. \ref{fig_complexity}(b), for each injection configuration, the corresponding drop response (D(t)) is photodetected, recorded by a 40 GHz oscilloscope over 10 $\mu s$ with 2 $ns$ sampling period, and mapped into a two-dimensional space generated upon the drop intensity $D(t)$ and its delayed version $D(t+\tau)$, with $\tau=10$ ns ($<\tau_{fc}=45$ ns \cite{borghi2021modeling}). In accordance with the Taken's theorem \cite{takens2006detecting} and its adaptation to a SCISSOR structure \cite{mancinelli2014chaotic}, this reconstruction creates an equivalent phase space where trajectories can be visualized. 
As the system evolves in time, its state, represented by a point ($D(t)$, $D(t+\tau)$), explores the reconstructed phase space, as more widely as the dynamics become more complex.  
Defining the phase space density $\rho$ as the fraction of the accessible phase space explored by the system, provides a quantitative estimate of the SCISSOR’s dynamical complexity. 
Figure \ref{fig_complexity}(c)(d) illustrates $\rho$ and the corresponding drop average optical power, respectively, as $\Delta\lambda$ varies from -500 pm to 500 pm with 20 pm resolution, and the optical power increases from 0.5 mW to 10 mW with 0.5 mW resolution (power estimated within the input waveguide, assuming 3 dB grating-coupler losses).
A few selected configurations highlighted with coloured circles in Fig. \ref{fig_complexity}(c) are linked to their respective drop time-trace in Fig. \ref{fig_complexity}(e).
Low $\rho$ corresponds to constant drop responses, achieved either when the optical power is too low ($<1$ mW) or when $\Delta\lambda$ is strongly detuned from the resonance, e.g. at $\Delta\lambda=280$ pm and 10 mW optical injection (purple case).
Reducing the detuning while maintaining the optical injection at 10 mW enables self-pulsation dynamics, initially in a single MRR ($\Delta\lambda=200$ pm, green case), then in two coupled MRRs ($\Delta\lambda=140$ pm, orange case), and eventually across all MRRs as $\Delta\lambda$ approaches the central resonance. Correspondingly, $\rho$ increases, reflecting increasingly complex dynamics. Regions of high density (large $\rho$, reddish region in Fig. \ref{fig_complexity} (c)) correspond to strong coupling between the three microrings where optical power is going back and forth between the MRRs that oscillate with aperiodic or chaotic dynamics. Regions of medium density (small $\rho$, light blue region in Fig. \ref{fig_complexity} (c)) correspond to a weak coupling between the three microrings where optical power is coupled to one or two MRRs and the dynamics is strongly nonlinear but not aperiodic. In fact, the highest $\rho$ obtained for $\Delta\lambda=20$ pm and 10 mW optical injection (light-blue case) coincides with a Fourier transform of the drop trace showing a loss of higher-order harmonics, a signature of chaotic dynamics. 
Under negative detuning the self-pulsation dynamics of a single MRR exhibits larger spike-to-quiescence contrast and refractory period \cite{donati2025all}.
In the SCISSOR under study, where the three resonances overlap, this is reflected in groups of three spikes interspaced by a refractory period, like those obtained at $\Delta\lambda=-100$ pm and 10 mW optical injection (blue case in Figs.\ref{fig_complexity}(c)(e)). The MRR closest to the input port spikes first, followed sequentially by the second and third, before the SCISSOR enters a global refractory period. 

When single MRRs enter self-pulsation dynamics the average drop optical power changes abruptly, hence providing a complementary marker for dynamical transitions. 
The mean optical power at the SCISSOR's drop port, shown in Fig. \ref{fig_complexity}(d), highlights clear discontinuities when transitioning from quiescence (purple case) to one self-pulsating MRR (green case), or from one to two self-pulsating MRRs (orange case). 
Tuning the optical injection parameters along the latter transition enables variable coupling interaction between two MRRs (see black circles 1 and 2 in Figs. \ref{fig_complexity}(d)(f)) supporting a wide range of optical and passive nonlinear transformations.
The transition from two to three self-pulsating MRRs is instead less apparent in drop power, but emerges as a discontinuity in $\rho$ (Fig. \ref{fig_complexity}(c)).  
Interestingly, as the system approaches the discontinuity in $\rho$ at negative detuning, the temporal dynamics become unstable, as observed for configurations 3, 4, and 5, as well as for the red operating point, in Fig. \ref{fig_complexity}(c), and confirmed by their Fourier spectra in Fig. \ref{fig_complexity}(e)(f). 
This shows that temporal instability can arise not only at high-$\rho$ operating points, but also in lower-power configurations exhibiting relatively low $\rho$ values when operating near specific discontinuities of the $\rho$ map (e.g., configuration 3).
In contrast, operating points located slightly away from this negative-detuning discontinuity remain temporally stable, with their Fourier spectra recovering well-defined harmonic peaks (configuration 6). This highlights that dynamical instabilities, as recorded over a 10 $\mu$s acquisition window under CW optical injection, are strongly localized at \cedge discontinuities even at low optical power, while neighboring regions support stable yet nonlinear dynamics.

\FloatBarrier
\section{Analogue/Digital Reservoir Computing}\label{sec_RC}

\begin{figure}[!t]
\centering
\includegraphics[width=0.8\textwidth]{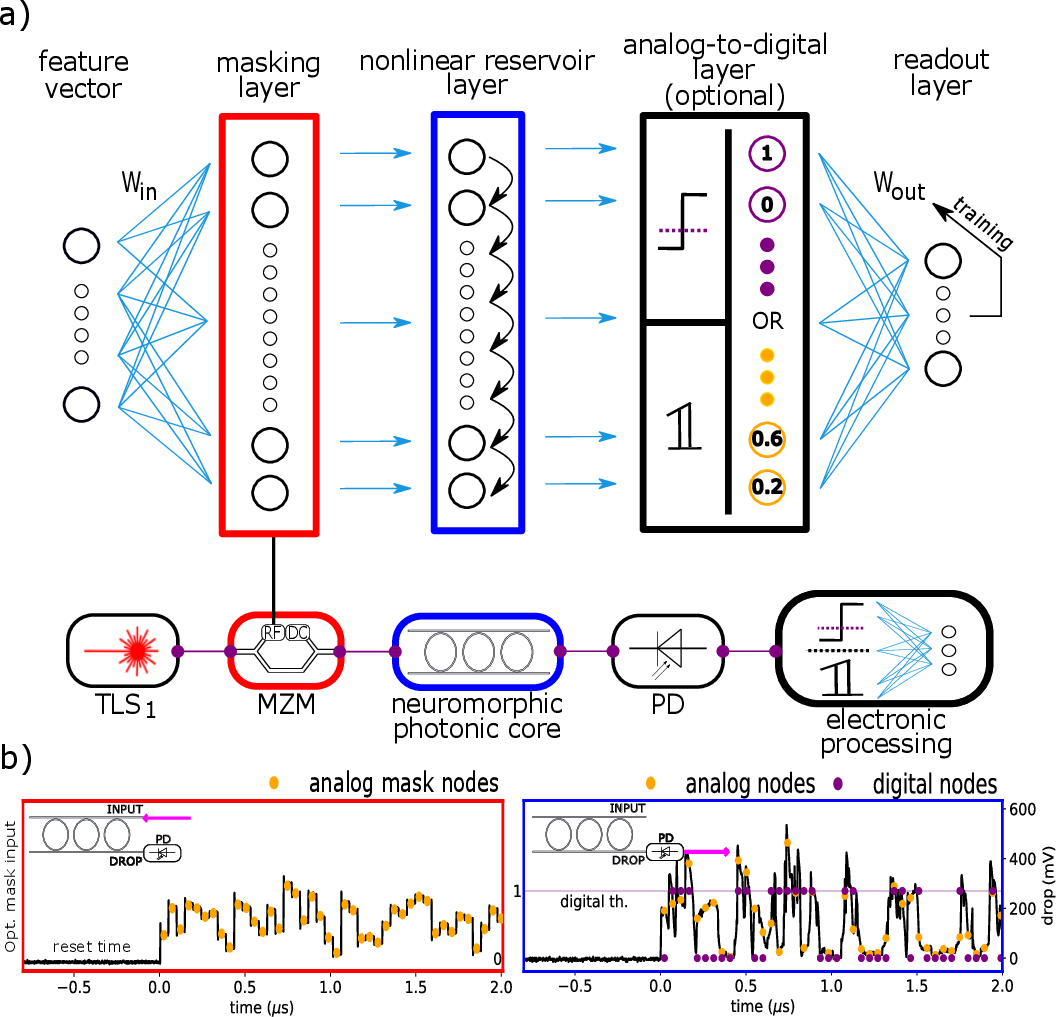}
\caption{(a) Analogue and digital RC scheme (top) and their photonic-electronic implementation (bottom). Nodes belonging to the masking layer (red) are temporally encoded using a Mach Zehnder modulator, while the reservoir state (blue) unfolds in the dynamical evolution of a physical SCISSOR neuromorphic core. The output layer (black) is computed electronically. For a complete setup description, see Experimental section. (b) Optical input signal encoding the masking layer nodes as amplitude values (left) and corresponding SCISSOR nonlinear response collected from the drop port (right), from which time-multiplexed reservoir nodes are extracted in analogue (yellow) or digital (purple) form according to an arbitrary digital threshold (th).}\label{RC_scheme}
\end{figure}

Here we investigate whether the rich variety of analogue and spiking nonlinear transformations supported by the SCISSOR device under test can be effectively exploited for information processing. 
The photonic-electronic RC architecture investigated in this work is sketched in Fig. \ref{RC_scheme}(a). Each datapoint is first linearly expanded offline into a masking layer composed of 150 nodes, obtained via different linear combinations of the original features. 
This number of nodes reflects a trade-off between reservoir size and experimental acquisition time, constrained by the waveform memory available for optical modulation (see \textit{Experimental setup} in Experimental section). 
The masking layer directly drives the RF port of a Mach-Zehnder modulator (MZM), generating a time-multiplex optical representation in which each neuron state is encoded as an amplitude level of a CW tunable laser source (see \textit{Optical encoding} in Experimental section, for encoding information). 
The modulated optical carrier is then injected into the SCISSOR neuromorphic core, where it undergoes a nonlinear optical transformation. 
A resting period of approximately 2 $\mu s$ ($>>\tau_{fc}, \tau_{th}$) is included, as also illustrated in Fig. \ref{RC_scheme}(b), to reset the SCISSOR internal memory before processing the next optical masking layer.
The modulation time per mask node is set to 48 ns, a value comparable to the free-carrier lifetime of the microring resonators ($\tau_{fc}=45$ ns \cite{borghi2021modeling}), ensuring that the SCISSOR operates continuously in a transient dynamical nonlinear regime while processing consecutive mask nodes \cite{appeltant2011information}. 
The SCISSOR's output signal is photodetected and received by an electronic processor, which samples it at a rate equal to the input mask modulation. In our implementation we use a 40 GHz oscilloscope to acquire the drop signal with 2 ns sampling period, and a computer for electronic post-processing.
As a result, the 150 analogue virtual nodes sampled at the electronic processor are temporally coupled through the SCISSOR’s intrinsic optical dynamics, forming a time-multiplexed reservoir layer of interacting nodes. The resulting reservoir states are then linearly combined in the electronic domain to compute the output layer. 
For edge computing applications, however, electronic processors may benefit from a binary representation of the reservoir state. For this reason, we compare the performance of standard analogue RC with a digital RC approach, inspired by \cite{owen2022ghz}, in which each virtual node assumes a value of 1 when its analogue state exceeds a digital threshold ($th$) and 0 otherwise (see Fig. \ref{RC_scheme}(b), right panel). The digital threshold is treated as a freely tunable electronic parameter and effectively acts as an additional Heaviside nonlinear operation applied to the reservoir state. When calibrated to detect optical spikes generated by the SCISSOR, this threshold directly exploits the intrinsic spiking dynamics of the device \cite{donati2025all}.

The SCISSOR-based RC system is benchmarked on the Iris Flower and Sonar classification tasks, two widely used datasets in the machine-learning community \cite{fisher1936use, gorman1988analysis}. Both tasks are evaluated across a wide range of optical injection conditions to identify which operating regime is the most effective for computation.
Specifically, the wavelength detuning $\Delta\lambda$ is swept from -320 pm to 320 pm in steps of 40 pm, while the injected optical power is varied in 0-10 mW with steps of 0.5 mW. 
The full dataset is partitioned into training, validation, and test subsets, with task-dependent proportions. The validation subset is used to optimize the regularization parameter ($\alpha$) of the logistic regression, thereby preventing overfitting.
For each injecting configuration, the SCISSOR’s dynamical response to the full dataset is acquired four times. 
Five-fold cross-validation is then applied to each acquisition to train the electronic output layer using logistic regression.
The classification accuracy ($Acc$) is defined as the fraction of correctly classified test datapoints. 
For each configuration, $Acc$ and its standard deviation $\sigma_{Acc}$ are computed as the mean and standard deviation over twenty scores (five cross-validation folds repeated across four independent acquisitions).
When performing digital RC, the entire procedure is repeated for multiple values of the digital threshold $th$. 

\FloatBarrier
\subsection{Iris Dataset classification task}\label{sec_Flower}

\begin{figure}[!t]
\centering
\includegraphics[width=0.9\textwidth]{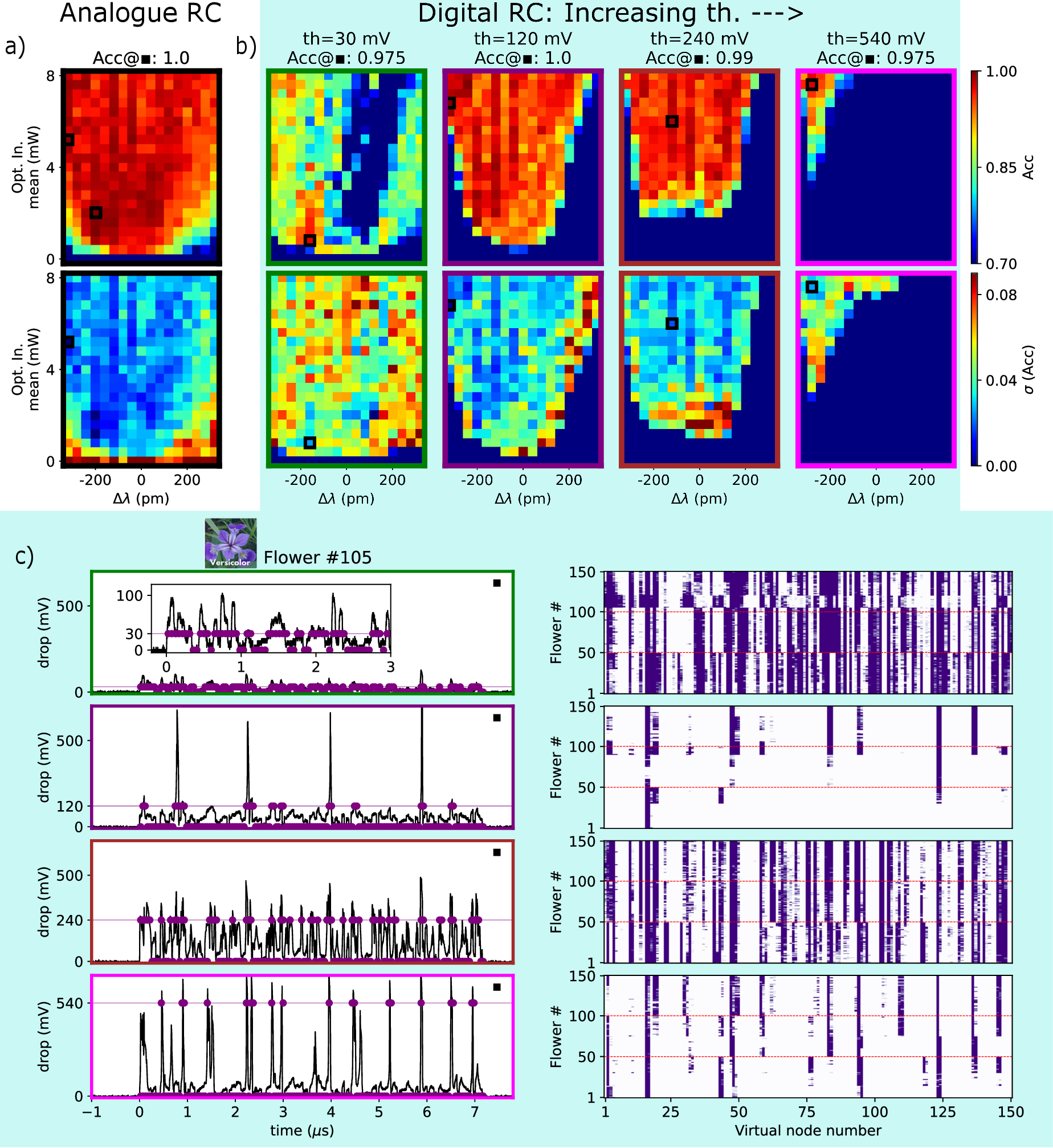}
\caption{Performance on the Iris task obtained using (a) analogue and (b) digital RC for increasing values of the digital threshold (th). Top panels report the classification accuracy ($Acc$), while bottom panels show the standard deviation of the accuracy ($\sigma_{Acc}$). (c) Best-performing digital RC configurations for increasing digital thresholds (black rectangles in (b)). Left panels show the optical nonlinear transformation produced by the SCISSOR for Iris sample $\#105$, while right panels report the corresponding digital reservoir state maps. Horizontal red lines indicate the boundaries between Setosa (1-50), Versicolor (51-100) and Virginica (101-150) classes. }\label{fig_Iris}
\end{figure}

The Iris dataset contains 150 datapoints from three Iris-flower species, Setosa, Versicolor, and Virginica, each described by four analogue features (sepal and petal lengths and widths). In our RC architecture (Fig. \ref{RC_scheme}(a)), this corresponds to a 4-150-150-3 topology for the input, mask, reservoir (analogue or digital), and output layers, respectively. 
For each cross-validation, the dataset is divided into 100 training, 20 validation, and 30 test datapoints.
Despite its widespread use, the Iris dataset represents a relatively simple classification task, already achieving $Acc \approx 99\%$ when logistic regression is applied directly to the masking layer (see \textit{Task error references} in Experimental section). It therefore serves as a useful benchmark to illustrate the operating principle of the proposed SCISSOR-based RC system and to relate its performance to previous works.

Figure \ref{fig_Iris}(a) shows the test accuracy (top panels) and the corresponding standard deviation (bottom panels) for analogue RC. 
The accuracy map exhibits a broad U-shaped region of optimal performance (darker red), indicating that effective nonlinear transformations of the masked optical input signal are achieved over a wide range of injection conditions. Notably, this optimal region extends to low input optical powers where, according to the characterization maps in Fig. \ref{fig_complexity}(c), the three MRRs do not exhibit self-pulsation and the dynamic is instead dominated by free-carrier-induced transients. Under these conditions, the system achieves $Acc=1$ in two optical injection configurations highlighted by black rectangles in Fig. \ref{fig_Iris}(a), thus increasing the performance achievable with a single silicon MRR \cite{borghi2021reservoir}.   

In digital RC, different reservoir states are generated from the same SCISSOR's dynamical trace by applying distinct digital thresholds in the electronic processor. Figure \ref{fig_Iris}(b) shows the performance obtained for four digital thresholds, each marked by a different coloured border.  
As the digital threshold increases (30 mV, 120 mV, 240 mV, 540 mV), a larger optical output is required to overcome the threshold and differentiate the digital virtual nodes states, forcing the optimal injection configurations to shift toward higher injection power. Configurations that fail to reach the threshold produce all-zero digital reservoir states for all flowers, causing the complete loss of differentiation and resulting in poor performance (dark blue regions).
For the highest thresholds investigated, the optimal configurations also shift toward more negative detuning where the SCISSOR, in agreement with its characterization shown in Fig. \ref{fig_complexity}(c), exhibits pronounced spiking dynamics. 
The optical responses and digital reservoir maps corresponding to the best performing configurations (highlighted by black rectangles in Fig. \ref{fig_Iris}(b)) are reported in Fig. \ref{fig_Iris}(c). 
At low thresholds and low injection optical power, the digital reservoir maps are dense, reflecting the low-amplitude nonlinear response of the passive device (green case).
As the injection power and threshold increase, the dynamics become increasingly spiking, leading to progressively sparser digital reservoir representations due to refractory effects that suppress consecutive virtual nodes following a spike (brown and pink cases).
Interestingly, an intermediate threshold value ($th=120$ mV) yields the sparsest digital representation that still achieves $Acc=1$. At this threshold, the electronic processor detects both well-defined output spikes and time intervals lying on the onset of spiking, producing enhanced variability in the digital patterns across the dataset. This balance between sparsity and variability enables the most effective digital representation for the Iris task we found. 
The crucial advantage of the SCISSOR-based digital RC, therefore, lies not in outperforming conventional analogue methods on this relatively simple task, but in achieving $Acc=1$ using extremely compact, sparse, and visually interpretable digital reservoir representations. As illustrated in Fig. \ref{fig_Iris}(c), Setosa, Versicolor and Virginica classes are separable even by an eye inspection of the digital maps, where they become represented by specific temporal nodes.

\begin{figure}[!t]
\centering
\includegraphics[width=0.8\textwidth]{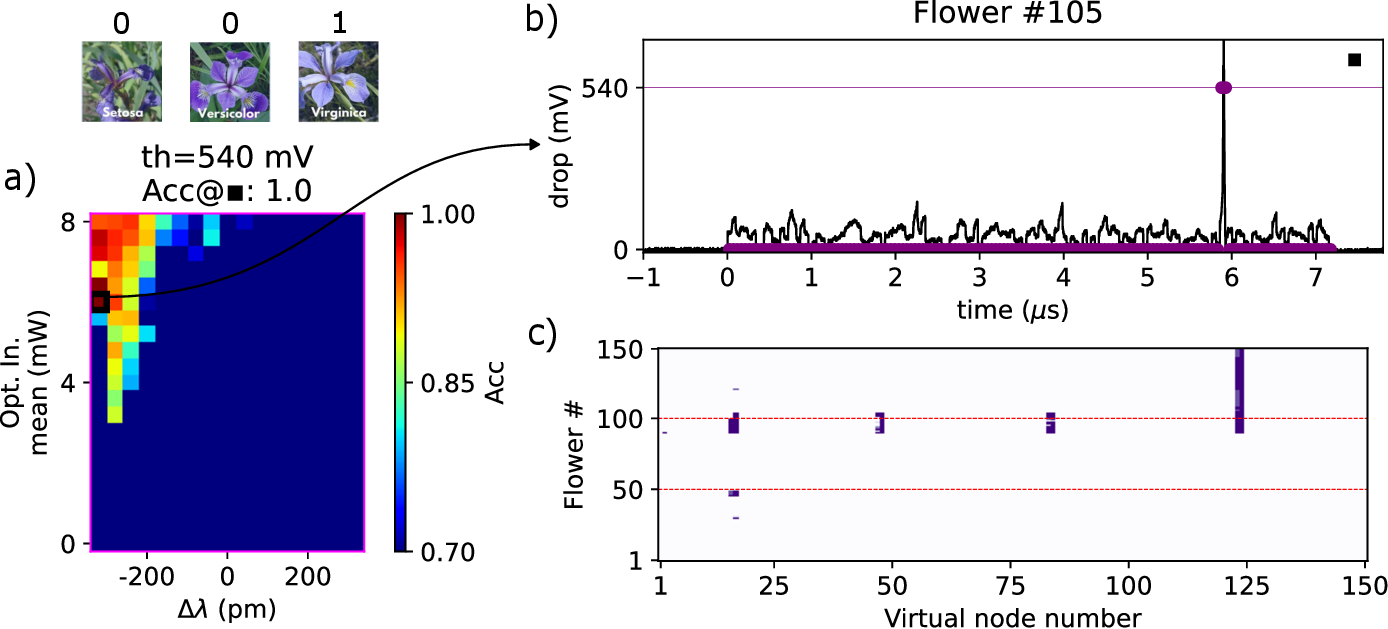}
\caption{Classification scores on the Iris task when the digital RC is trained to recognize only the Virginica specie. (a) Accuracy ($Acc$) scores obtained when setting a digital threshold (th) of 540 mV. Under an optical injection of $\Delta\lambda=-320$ pm and 5.85 mW input power (within the input waveguide) the SCISSOR obtained accuracy 1. (b) The corresponding SCISSOR optical nonlinear transformation of Flower $\#105$ (belonging to the Virginica class), and (c) map of the digital patterns obtained for the full dataset, show that only one spike allows solving the task. Horizontal red lines in the digital map indicate the boundaries between Setosa (1-50), Versicolor (51-100) and Virginica (101-150) classes.} \label{fig_Iris_Class3}
\end{figure}

Finally, the flexibility of the SCISSOR nonlinear dynamics can be exploited for more specific task-optimizations. As an illustrative example, Fig. \ref{fig_Iris_Class3} shows the performance obtained when the SCISSOR-based digital RC is trained to recognize only the Virginica class against the other two species. 
In this case, combining an injection configuration of $\Delta\lambda=-320$ pm and 5.85 mW mean input optical power (black rectangle in Fig. \ref{fig_Iris_Class3}(a)), with a digital threshold $th=540$ mV in the electronic processor, allows the task to be solved ($Acc=1$) predominantly by the activation of just a single temporal node, highlighting the potential of the SCISSOR-based approach for highly efficient and fast electronic post-processing.  
By comparison, a digital reservoir computer based on a VCSEL spiking neuron using a similar encoding scheme achieves on the Iris task 91.7$\%$ accuracy using 512 reservoir virtual nodes and exceeds $97\%$ only when scaled to 1024 virtual nodes \cite{owen2022ghz}, whereas the present system achieves error-free operation using only 150 virtual nodes and sparse activations.  

\FloatBarrier
\subsection{Sonar Dataset classification task}\label{sec_R_vs_M}

\begin{figure}[!t]
\centering
\includegraphics[width=0.9\textwidth]{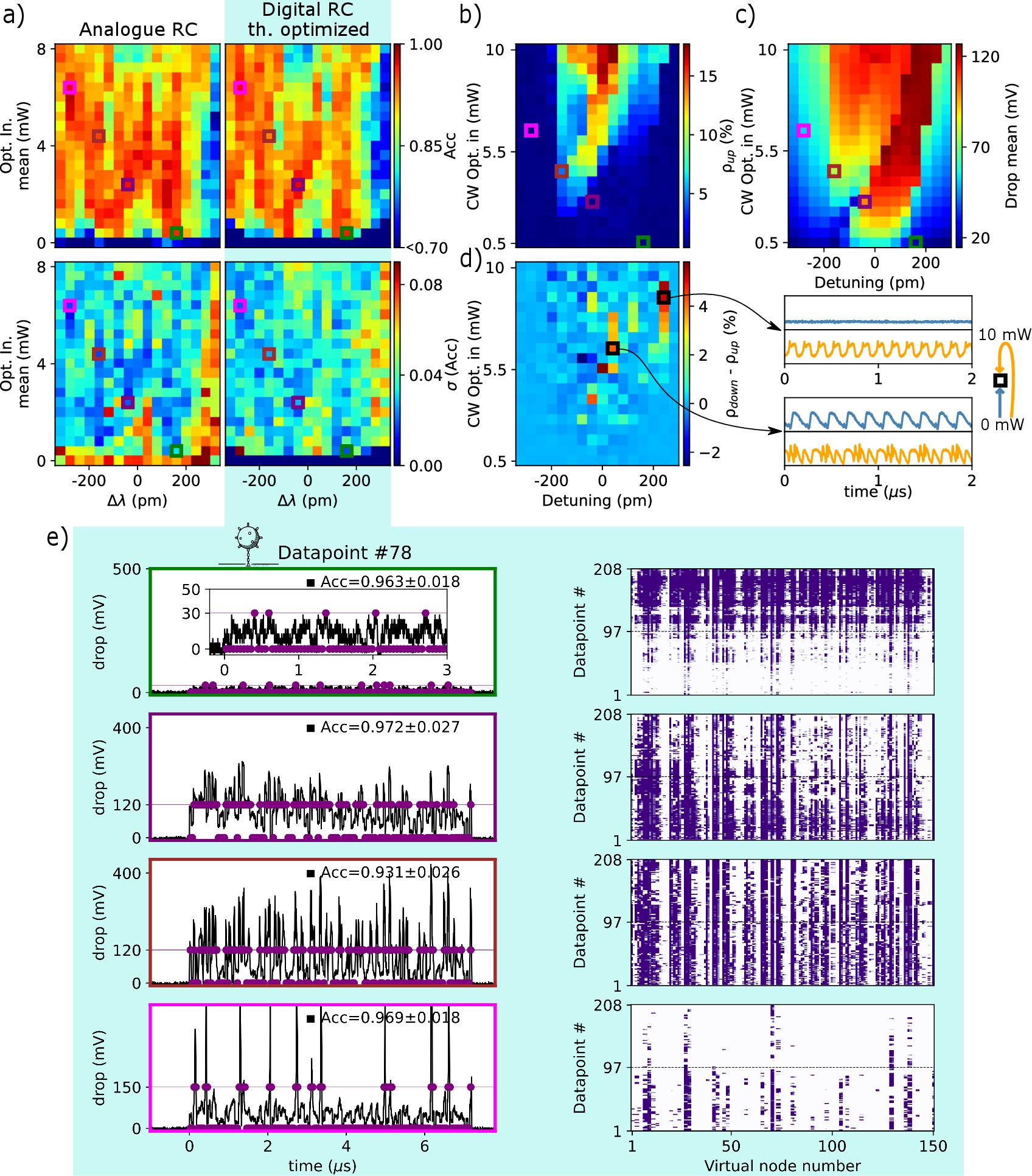}
\caption{Sonar task performance and its connection to the SCISSOR nonlinear dynamics. (a) Analogue (left panels) and digital (right panels) RC performance. For digital RC, the threshold (th) is independently optimized for each combination of $\Delta\lambda$ and input optical power. Top and bottom panels report the classification accuracy ($Acc$) and its standard deviation ($\sigma_{Acc}$), respectively. (b) Nonlinear dynamical complexity of the SCISSOR output quantified by $\rho_{up}$. (c) Mean value of the optical drop signal. (d) Spiking bistability map defined as $\rho_{down}$-$\rho_{up}$, and, on the right, examples of dynamics achieved in a target configuration (black rectangle) by tuning the optical power either directly from 0 mW or from 10 mW. (e) Four selected digital RC operating points, highlighted by coloured rectangles in panels (a-c). Left panels show the SCISSOR optical nonlinear transformation for the Sonar sample $\#78$, while right panels display the corresponding digital reservoir state maps. Horizontal red lines in the digital map indicate the boundaries between Rock (1-97) and Mine (98-208) classes.}
\label{fig_Sonar}
\end{figure}

The Sonar dataset contains 208 datapoints that need to be distinguished as either rocks (samples 1-97) or mines (samples 98-208). Each datapoint is described by 60 analogue features \cite{gorman1988analysis}. Accordingly, the RC architecture shown in Fig. \ref{RC_scheme}(a) assumes a 60-150-150-2 configuration for this task, corresponding to the feature, masking, reservoir (analogue or digital) and output layers, respectively.
For each cross-validation, the dataset is divided into 130 training, 20 validation, and 58 test datapoints.
Although this dataset has only two output classes, one fewer than the Iris task, it reaches an accuracy of only $\approx 80\%$ when logistic regression is applied directly to the masking layer (see \textit{Task error references} in Experimental section). The Sonar classification problem is therefore significantly more challenging, providing a more stringent benchmark for evaluating the nonlinear transformations generated by the SCISSOR-based RC system.

The performance obtained in analogue and digital RC are shown in Fig. \ref{fig_Sonar}(a). 
For digital RC, the results shown correspond to a digital threshold that has been optimized independently for each injection configuration.
This representation provides a compact alternative to the different threshold-maps presented for the Iris task in Fig. \ref{fig_Iris}(b). 
Under this representation, clear similarities emerge between analogue and digital RC. 
In both cases, the performance map broadly resembles that obtained for the Iris dataset (Fig. \ref{fig_Iris}(a)). However, the higher intrinsic complexity of the Sonar task causes the best-performing configurations to emerge more distinctly, yielding sharper performance contrasts across the injection-parameter space. These contrasts closely mirror features observed in the SCISSOR characterization maps shown in Fig. \ref{fig_complexity}(c)(d).
To facilitate a direct comparison, the characterization maps of the drop complexity $\rho$ and mean drop power in Fig. \ref{fig_complexity}(c)(d) were resampled from a 20 pm detuning resolution to match the 40 pm detuning resolution used during the task experiments. The resampled maps are reported in Fig. \ref{fig_Sonar}(b)(c), respectively.
Notably, the regions of highest performance accuracy and lowest standard deviation in the Sonar task (Fig. \ref{fig_Sonar}(a)) coincide with injection conditions that place the SCISSOR at the \cedge between two or three self-pulsating MRRs, as shown in Fig. \ref{fig_Sonar}(b)(c).  
These configurations correspond to intermediate values of the complexity $\rho$, where the nonlinear dynamics are rich but remain well structured. 
In contrast, the region of highest complexity $\rho$ in Fig. \ref{fig_Sonar}(b), where all three MRRs are deeply nonlinearly coupled and the fft of the drop signal loses distinct harmonic peaks (Fig. \ref{fig_complexity}(e), light-blue case), corresponds to a marked degradation of performance in Fig. \ref{fig_Sonar}(a). 
This correlation confirms that nonlinear distortions arising at the \cedge of multiple MRRs support effective reservoir transformation for this task, yielding a maximum analogue RC accuracy of $0.984 \pm 0.011$. 

Beyond accuracy, the sparsity of the digital reservoir states varies significantly across different injection configurations. To illustrate this effect, four representative operating points are highlighted by coloured rectangles in Fig. \ref{fig_Sonar}(a-c), and their SCISSOR nonlinear transformations to a representative datapoint (sample $\#$78) are shown in Fig. \ref{fig_Sonar}(e). 
In general, \cedge configurations located at higher optical power and negative detuning exhibit progressively stronger spiking dynamics. 
Remarkably, despite the digital reservoir states becoming increasingly sparse as the spiking dynamics dominate, the classification accuracy remains high. As shown in Fig. \ref{fig_Sonar}(e) (left panels), the most sparse configuration (pink) still achieves an accuracy of $0.969 \pm 0.018$. 
Notably, these results surpass those obtained using logistic regression applied directly to either the feature or masking layers (see \textit{Task error references} in Experimental Section), as well as previously reported performance based on back-propagated feed forward neural networks with sigmoid activation function (accuracy up to 0.904) \cite{gorman1988analysis}, and support vector machines (accuracy up to $\sim 0.92$) \cite{salankar2007svm}. 

\begin{figure}[!t]
\centering
\includegraphics[width=1\textwidth]{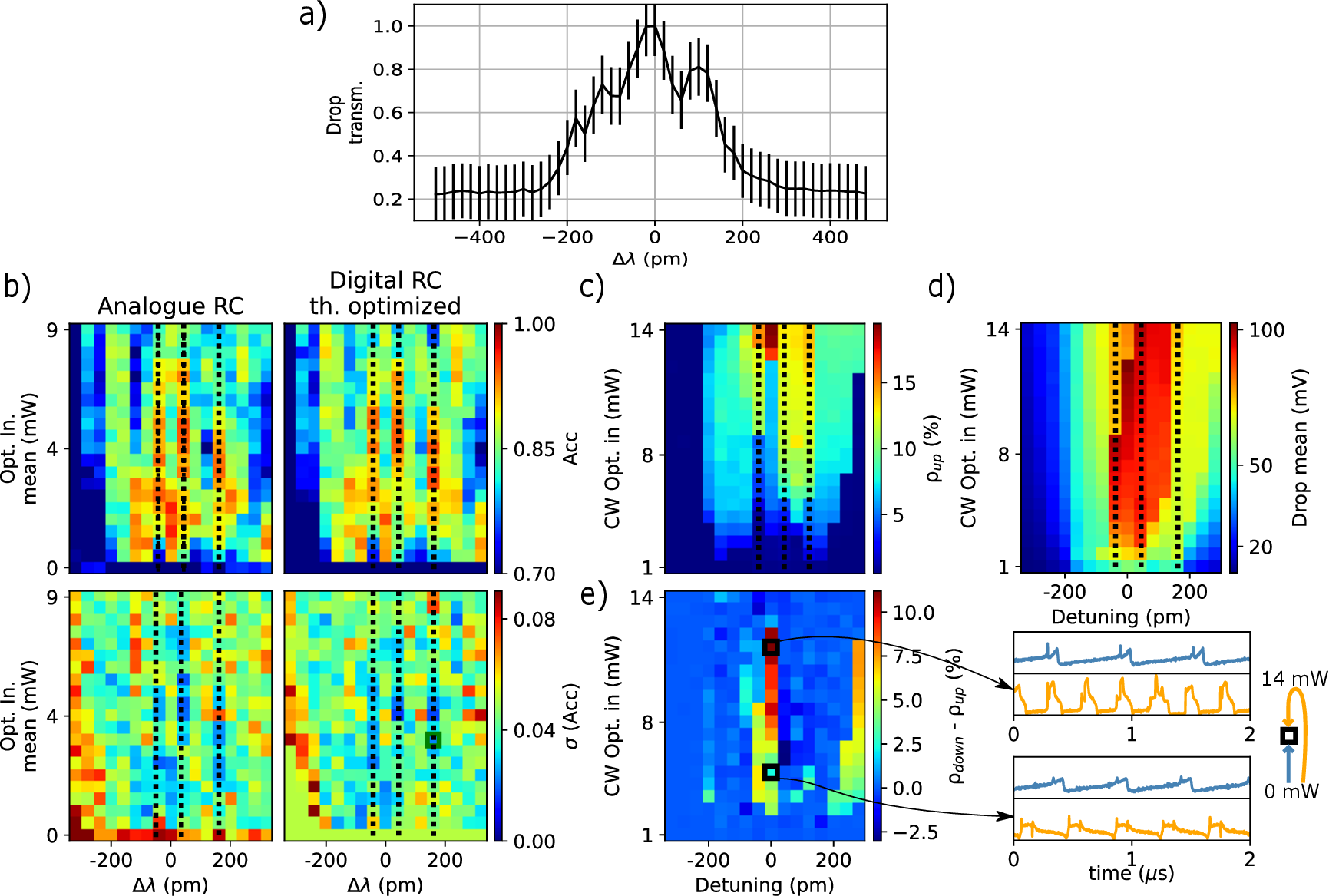}
\caption{Sonar task performance and its connection to the SCISSOR nonlinear dynamics for a new 3-MRRs SCISSOR. (a) Linear drop transmission. (b) Analogue (left panels) and digital (right panels) RC performance. For digital RC, the threshold (th) is independently optimized for each combination of $\Delta\lambda$ and input optical power. Top and bottom panels report the classification accuracy ($Acc$) and its standard deviation ($\sigma_{Acc}$), respectively, computed over 5-fold cross-validation and averaged over four independent acquisitions. (c) Nonlinear dynamical complexity of the SCISSOR output quantified by $\rho_{up}$. (d) Mean value of the optical drop signal. (e) Spiking bistability map defined as $\rho_{down}$-$\rho_{up}$, and, on the right, examples of dynamics achieved in a target configuration (black rectangle) by tuning the optical power either directly from 0 mW or from 14 mW. Vertical black dashed lines help tracking the \cedge condition between task and CW characterization maps. }
\label{fig_Sonar_2}
\end{figure} 

Among the four highlighted configurations in Fig. \ref{fig_Sonar}(a), the operating point at $\Delta\lambda=160$ pm and 0.4 mW optical injection (green) was intentionally selected to lie out from any \cedge. Instead, it is positioned just below the onset of a thermal bistability, as indicated by the first jump in the mean drop optical power in Fig. \ref{fig_Sonar}(c) and the corresponding minimum-$\rho$ region in Fig. \ref{fig_Sonar}(b). Despite remaining quiescence, this configuration exploits the proximity to the thermal bistable regime during the task, achieving a high accuracy of $0.963 \pm 0.018$. 
The associated digital reservoir map (Fig. \ref{fig_Sonar}(e), top-right panel), reflects this thermal bistable behaviour, producing dense and sparse activation regions for datapoints belonging to the two classes, respectively.  
This observation suggests that optical (quiescent) bistabilities, may represent a complementary mechanism for optical computing, capable of enhancing class discrimination.  

Finally, the SCISSOR also exhibits spiking bistabilities, as characterized in Fig. \ref{fig_Sonar}(d). Here, $\rho_{up}$ and $\rho_{down}$ denote the complexity of the drop signal measured when the CW optical injection power is increased directly from zero ($\rho_{up}$) or decreased from a high-power state (10 mW) toward the target operating condition, respectively. Large differences between $\rho_{up}$ and $\rho_{down}$ identify regions of spiking bistabilities. 
The configuration at $\Delta\lambda=240$ pm (black rectangle in Fig. \ref{fig_Sonar}(d)) corresponds to the spiking bistability in a single MRR switching from quiescence to self-pulsation dynamics. In contrast, the configuration at $\Delta\lambda=40$ pm highlights a region where the coupling among all MRRs causes bistable switching between two non-quiescent, temporally periodic states. 
During a task, the amplitude modulations imposed by the masking layer can transiently trigger such switching, increasing the effective complexity of the SCISSOR's response \cite{donati2024spiking}.
Consistently, the central spiking bistable region in Fig. \ref{fig_Sonar}(d) aligns with a high-accuracy and low $\sigma(Acc)$ region in Fig. \ref{fig_Sonar}(a).
This enhancement is absent for isolated single-MRR spiking bistability at $\Delta\lambda=240$ pm. 

To further consolidate the \cedge computing concept, the Sonar task was also experimentally tested using a second SCISSOR core composed of three nominally identical MRRs (radius of 6.75 $\mu m$, center-to-center distance 63.617 $\mu m$, symmetric bus-waveguide gap of 237 nm). Due to fabrication defects, the linear drop transmission in Fig. \ref{fig_Sonar_2}(a) exhibits slightly-shifted MRR resonances, which nevertheless retain partial spectral overlap. 
For this SCISSOR, the wavelength detuning $\Delta\lambda$ is defined relative to an arbitrary chosen central wavelength $\lambda_0=1553.575$ nm.
Figure \ref{fig_Sonar_2}(b) shows the performance in analogue and digital RC obtained with this SCISSOR on the Sonar task, while Fig. \ref{fig_Sonar_2}(c)(d) show its CW characterization. 
In particular, Fig. \ref{fig_Sonar_2}(c)(d) reveals vertical \cedge conditions at $\Delta\lambda=\pm 40$ pm, and $\Delta\lambda=160$ pm, highlighted by vertical black dashed lines, where either the drop complexity and/or the drop average optical power undergo a sudden change.
Correspondingly, the best performing analogue and digital RC configurations in Fig. \ref{fig_Sonar_2}(b) emerge at these same detuning values, supporting the reproducibility of the \cedge operating principle across different devices.
In addition, $\Delta\lambda=0$ pm identifies a regime of strong spiking bistability within the SCISSOR (see Fig. \ref{fig_Sonar_2}(e)), whose low-power operation correlates with enhanced accuracy and reduced standard deviation in Fig. \ref{fig_Sonar_2}(b)).

\FloatBarrier
\section{Conclusions and future perspectives}\label{sec_Conclusions}

This work demonstrates that scaling a single MRR neuron into three coupled MRRs arranged in a SCISSOR geometry provides a compact (150$\times$20 $\mu m^2$) and passive nonlinear core for integrated optical computing applications. 
Beyond its excellent performance metrics in both analogue and digital (spiking) reservoir computing approaches obtained in the Iris and Sonar dataset tasks, a central outcome of this work is the identification of \cedge operating regimes coinciding with discontinuities in dynamical complexity (estimated by the Taken's theorem) and/or mean optical power at the boundary of strong and weak coupling regimes. These regimes, for the more challenging Sonar task, support high and robust accuracy.
Low-power optical spiking bistabilities also contribute to enriching the temporal structure of the reservoir response leading to improved computing capabilities. 
Together, these findings suggest a physics-driven criterion for identifying favourable operating conditions, significantly reducing the need for exhaustive parameter searches in future works and may also serve as central points around which engineering training algorithms on-chip.

The average input optical power that is needed to activate free-carrier and thermal nonlinearities within the SCISSOR core depends on the optical detuning $\Delta\lambda$. In particular, less than 8 mW are required for the most demanding regimes exhibiting strongly coupled spiking dynamics. These spiking regimes are currently limited to megahertz speeds by the thermal lifetime of the MRRs; however, higher-speed operation may be achieved using graphene-on-silicon MRRs \cite{jha2022photonic} or embedding, for instance, the MRR waveguides within a p-i-n junction \cite{shetewy2024demonstration}. This in turn would increase the processing speed for RC. Alternatively, the input mask can also be optimized in future works, such to maximize the diversification of the SCISSOR output responses and digital patterns using less output virtual nodes.
The computational richness of the SCISSOR structure is further expandable by using multiple detuned injection wavelengths, which would introduce parallel and interacting feedback pathways within the structure and enable wavelength multiplexed fan-out capabilities, or by tuning the inter-resonator coupling via integrated heaters or p-n junctions (which would also allow tuning the \cedge regions), while remaining compatible with MRR-weight bank schemes and electrically driven spiking approaches \cite{xiang2025photonic}.
Overall, this work motivates the scaling of architectures comprising multiple SCISSOR cores operated at their respective \textit{coupling-edges}, either in parallel or coupled via tuneable connections (e.g., via phase shifters \cite{lugnan2025emergent}), as a compelling research direction to be explored in future works.

\FloatBarrier
\section{Experimental section} \label{sec_Experimental}

\threesubsection{Experimental setup} 

\begin{figure}[!t]
\centering
\includegraphics[width=1\textwidth]{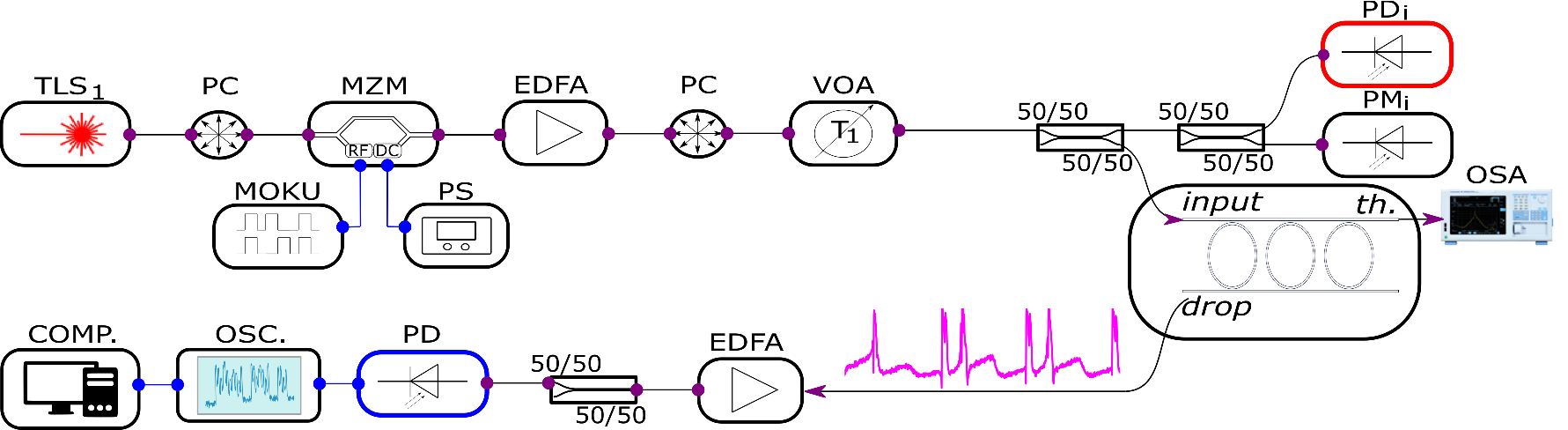}
\caption{Experimental setup for the CW characterization of the SCISSOR and the optical encoding of the computing task to be processed. Symbols are defined in the main text.  }\label{fig_Setup}
\end{figure}

In the experimental setup, light generated by a tunable laser source is amplitude-modulated by a Mach–Zehnder modulator (MZM) to sequentially encode the node states of the masking layer. The optical power of the modulated signal is adjusted using an erbium-doped fibre amplifier (EDFA), followed by a variable optical attenuator (VOA) employed for automated tuning and calibration routines.
The optical signal is then split into two equal (50/50) arms. One arm is used as a monitoring path, where the input signal waveform and its average optical power are recorded using a photodetected (PD, connected to a 40 GHz oscilloscope) and a power meter (PM), respectively. The second arm is coupled to the input port of the SCISSOR structure under test.
To maintain a stable detuning $\Delta\lambda$, the chip temperature is actively regulated using a proportional-integral-derivative (PID) controller connected to a Peltier cell and a 10 k\si{\ohm} thermistor. This limits resonance wavelength fluctuations to within 3 pm over a five-hour monitoring period, comparable to the 2 pm wavelength stability of the input lasers \cite{donati2025all}.
A fibre array is used to inject the optical signal into the chip and to collect both the through-port and drop-port responses of the SCISSOR. The through signal is monitored using an optical spectrum analyser (OSA) for coupling and wavelength-alignment calibration. The drop signal, which is used for the reservoir computing implementation, is amplified by an EDFA, photodetected, and subsequently acquired by an oscilloscope (OSC) before being transmitted to a computer (COMP) for electronic post-processing.

The oscilloscope is operated with a 40 GHz bandwidth and a sampling period of 2 ns. 
Each mask node, amplitude-encoded in the optical domain for 48 ns, relates to 24 output samples ($s_i$, $i=1\dots 24$) acquired in the corresponding time window. 
In analogue RC, the state of each virtual node is defined by the value of the 12th sample.
In digital RC, all 24 samples are examined, and a digital value of 1 is assigned if the condition $s_i>th$ is satisfied for at least one sample $i$.

\threesubsection{Optical encoding} 

\begin{figure}[!t]
\centering
\includegraphics[width=0.6\textwidth]{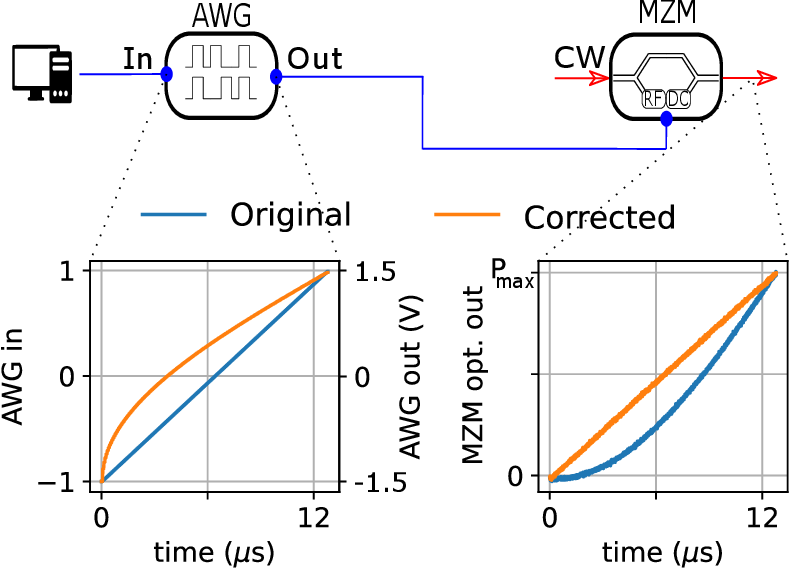}
\caption{Compensation of the Mach–Zehnder modulator (MZM) nonlinear transfer function to ensure a linear encoding of the masking signal in the optical domain.}\label{fig_Corr}
\end{figure}

Undesired static nonlinear distortions arising from the intrinsic cosine-squared intensity transfer function of the Mach–Zehnder modulator (MZM) must be compensated in order to isolate the nonlinear response of the SCISSOR under study \cite{bazzanella2022microring}. To this end, a dedicated calibration procedure is performed prior to each task experiment, as follows.
A linearly increasing voltage waveform (left panel, blue curve) is generated by the arbitrary waveform generator (AWG) and applied to the MZM using the maximum available modulation voltage range (3 V peak-to-peak).
The MZM bias voltage is then adjusted such that the minimum of the transfer function is reached at the maximum modulation amplitude (right panel, blue curve). This operating point is crucial, as the encoding scheme employed during the task includes a reset interval between consecutive masking layers, during which the optical power entering the SCISSOR is minimized to allow the internal nonlinear dynamics to decay (see Fig. \ref{RC_scheme}).
Under these conditions, the MZM maps the linearly increasing electrical waveform onto a nonlinear optical intensity response that reflects the curvature of its transfer function (right panel, blue curve). Although static, this distortion is non-negligible, as it introduces unintended nonlinear preprocessing of the input signal. A correction function is therefore estimated and applied directly to the AWG waveform (left panel, orange curve), enabling the recovery of a linear optical waveform at the MZM output (right panel, orange curve). Whenever the bias voltage applied to the MZM or the AWG voltage range is changed, a different portion of nonlinear transfer function is accessed and the correction function must be re-estimated. 
The MZM-correction phase is performed using the setup's input line described in Experimental section (\textit{task error references}), with the EDFA bypassed. 
The correction function is applied directly to the electrical waveforms corresponding to each masking layer prior to optical encoding, ensuring their linear reproduction in the optical domain.

\threesubsection{Task error references} 

\begin{figure}[!t]
\centering
\includegraphics[width=1\textwidth]{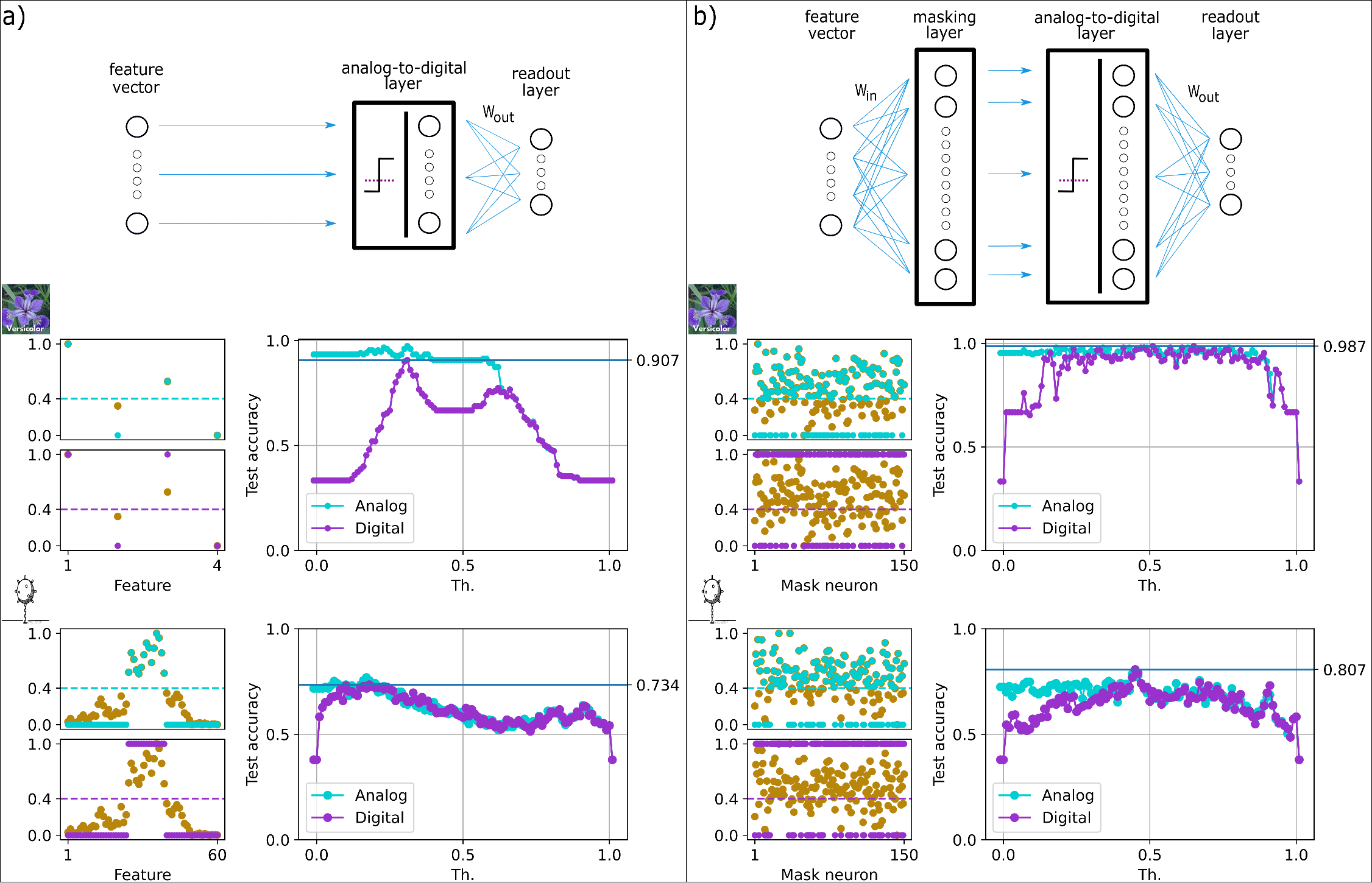}
\caption{Reference test accuracy scores for the Iris and Sonar classification tasks. A thresholding layer is applied directly to (a) the raw feature layer and (b) the masking layer, respectively. Logistic regression is used to train and compute the output layer. The SCISSOR nonlinearity, which would otherwise couple the reservoir nodes, is completely bypassed. }\label{fig_Input_Ref}
\end{figure}

The reference scores for the Iris and Sonar classification tasks are obtained by completely bypassing the SCISSOR layer, and therefore its nonlinear transformation, while retaining the thresholding operation.

In the raw-feature reference case, the threshold layer is applied directly to the feature layer, after which logistic regression is used to compute the output layer. Each set of data entering the feature layer is independently normalized in the range 0-1, according to its maximum and minimum values.
Then, the digitalization scheme adopted in the main manuscript is preserved: nodes above threshold assume a value of 1 and 0 otherwise. For consistency, this procedure is extended to the analogue case, where nodes above threshold retain their analogue value, while nodes below threshold are set to zero.
The test accuracies obtained on the Iris and Sonar datasets are shown in Fig. \ref{fig_Input_Ref}(a) as a function of the threshold value (th). 
For $th=0$, all information is preserved in the analogue case, resulting in high accuracy, while in the digital case the input collapses to an all-ones vector, yielding minimal performance. 
As $th$ increases, an increasing number of nodes are set to zero. In the analogue case this progressively removes information, whereas in the digital case it introduces differentiation between patterns, leading to improved performance. Further increasing $th$ eventually drives both analogue and digital representations toward all-zero vectors, causing a degradation in accuracy in both cases.

We then repeated the same procedure after expanding the raw feature layer into a 150-node masking layer. Here, the normalization is applied directly to the masking layer (and not to the original feature layer), with each masking node normalized to the range 0–1 based on the minimum and maximum values of the layer, for each datapoint.
The corresponding test accuracies, shown in Fig. \ref{fig_Input_Ref}(b), are significantly more robust with respect to the threshold parameter compared to the raw-feature case and reach higher maximum scores, namely 0.987 for the Iris dataset and 0.807 for the Sonar dataset in the digital case.
Finally, the classification tasks were also solved by applying logistic regression directly to the optically encoded masking layer (acquired by $PD_i$), which was normalized using the same procedure. 
In this case, each input mask node state is the average of the 24 oscilloscope samples acquired in the allocated time window. Without applying any threshold, therefore preserving the full analogue information, we obtain average accuracies of 0.990 for the Iris dataset and 0.776 for the Sonar dataset. These values closely match those reported in Fig. \ref{fig_Input_Ref}(b) for the analogue case at $th=0$, confirming the quality of the linear optical encoding discussed (see \textit{Optical encoding} in Experimental section).

\medskip

\medskip
\textbf{Acknowledgements} \par 
We gratefully thank Dr. Alessio Lugnan for useful suggestions and interesting discussions. The authors also acknowledge funding support from the UKRI Turing AI Acceleration Fellowships Programme (reference EP/V025198/1), EU Pathfinder Open project ‘SpikePro’ (Grant ID 101129904), the UK Multidisciplinary Centre for Neuromorphic Computing (reference UKRI982), the UK Neuromorphic Computing Hardware Semiconductor Innovation Knowledge Centre (IKC), and the European Research Council (ERC) under the European Union’s Horizon 2020 research and innovation programme (Grant Agreement No. 788793, BACKUP), which has now ended, and during which the chip used in these experiments was fabricated.

\medskip
\textbf{Data availability} \par 
Data underlying the results presented in this paper are available in Ref. \cite{Donati_data_link}.

\medskip

\printbibliography

\end{document}